\documentclass[twocolumn,aps,prb,10pt,amssymb,floatfix,superscriptaddress]{revtex4-1}

\usepackage{hyperref}
\usepackage{graphicx}
\usepackage{amsmath,amsfonts,amsthm,bm}
\hypersetup{colorlinks=true, linkcolor=blue, citecolor=blue, urlcolor=blue}

\begin{document}

\date{\today}

\author{Hugo Tschirhart}
\email{hugo.tschirhart@uni.lu}
\affiliation{Physics and Materials Science Research Unit, University of Luxembourg, 1511 Luxembourg, Luxembourg}

\author{Ernest Teng Siang Ong}
\affiliation{School of Physical and Mathematical Sciences, Nanyang Technological University, 21 Nanyang Link, Singapore 637371}

\author{Pinaki Sengupta}
\affiliation{School of Physical and Mathematical Sciences, Nanyang Technological University, 21 Nanyang Link, Singapore 637371}

\author{Thomas L. Schmidt}
\affiliation{Physics and Materials Science Research Unit, University of Luxembourg, 1511 Luxembourg, Luxembourg}

\title{Phase diagram of spin-$1$ chains with Dzyaloshinskii-Moriya interaction}

\begin{abstract}

We investigate an antiferromagnetic spin-$1$ Heisenberg chain in the presence of Dyzaloshinskii-Moriya interactions (DMI) and an external magnetic field. We study the resulting spin chain using a combination of numerical and analytical techniques. Using DMRG simulations to determine the spectral gap and the entanglement spectrum, we map out the phase diagram as a function of magnetic field strength and DMI strength. We provide a qualitative interpretation for these numerical findings by mapping the spin-$1$ chain on a spin-$1/2$ ladder and using a bosonization approach.

\end{abstract}
\maketitle
\section{Introduction}
\label{intro}

Haldane's work\cite{haldane1,haldane2} on quantum spin-$1$ chains has led to one of the first examples of a symmetry-protected topological phase. Indeed, the conjectured existence of a finite gap between the ground state and the excited states in antiferromagnetic spin-$S$ Heisenberg chains with integer $S$ has sparked a substantial research activity in integer-spin chains, mostly by means of numerical studies based on Monte Carlo simulations\cite{nightingale_montecarlo,albuerque_spin1_montecarlo,kim_spin1_montecarlo,
matsumoto_spin1_montecarlo,pinaki_spin1_montecarlo1,
pinaki_spin1_montecarlo2,pinaki_spin1_montecarlo3} or the Density Matrix Renormalization Group (DMRG).\cite{white_spin1_dmrg1,white_spin1_dmrg2,hu_spin1_dmrg,
moukouri_spin1_dmrg1,moukouri_spin1_dmrg2} It has also motivated experimentalists to search for host materials for spin-$1$ chains, and those were eventually realized in $\mathrm{Ag VP_2S_6}$ \cite{mutka_spin1_magnet,asano_spin1_magnet,takigawa_spin1_magnet1,
takigawa_spin1_magnet2}, $\mathrm{SrNi_2V_2O_8}$, \cite{zheludev_spin1_magnet,pahari_spin1_magnet,bera_spin1_magnet} or $\mathrm{CsNiCl_3}$.\cite{buyers_experimental_evidence}

This Haldane phase constitutes a topological phase, so its ground state degeneracy depends on the applied boundary conditions. In the case of a spin-$1$ chain with open boundary conditions, two uncoupled, localized spin-$1/2$ states appear at the ends of the chain which lead to a fourfold degenerate ground state. In contrast, the ground state for periodic boundary conditions is non-degenerate. Applying a constant magnetic field leads first to a closing of the gap. Upon further increase of the field, the gap reopens and the system enters a (topologically trivial) ferromagnetic phase.

However, as the Haldane phase constitutes only a symmetry-protected topological phase, these ground state degeneracies and the spectral gap are not robust to all perturbations.\cite{pollman_entropy_spectrum} A helical magnetic field, for instance, which is unitarily equivalent to Dzyaloshinskii-Moriya interaction (DMI),\cite{dzyaloshinskii,moriya} destroys the topological phase. In particular, the extreme case of a staggered magnetic field was studied in detail in Refs.~[\onlinecite{Tsukano-staggered1}] and [\onlinecite{Tsukano-staggered2}]. It was found that the spectrum in this case remains gapped for arbitrary magnetic field strengths, without any phase transition between the Haldane phase and the antiferromagnetically ordered phase.

The purpose of the present work is a detailed analysis of the crossover between these two extreme cases. We will therefore investigate a spin-$1$ chain in the presence of a helical magnetic field with a given pitch angle $\varphi$, where $\varphi = 0$ corresponds to the case of constant magnetic field, whereas $\varphi = \pi$ corresponds to the limit of a staggered field, i.e., a field with opposite orientation on neighboring lattice sites. As we show below, a spin chain in a helical magnetic field can be mapped onto a spin chain in a constant field but with DMI. Spin systems with DMI have recently attracted significant attention, for instance in the study of magnetized thin films,\cite{hrabec_dmi_thinfilms,heide_dmi_thinfilms,rohart_dmi_thinfilms}, graphene, \cite{ajejas_dmi_graphene} spin-$1/2$ ladders,\cite{ueda_dmi_spinladder} multiferroic phases, \cite{sergienko_dmi_montercarlo}, anisotropic Heisenberg antiferromagnets,\cite{parente_dmi_meanfield} or XY spin chains.\cite{farajollapour_dmi_renormalizationgroup,
gombar_dmi_renormalizationgroup}

On the numerical side, we have studied the ground state properties of the $S=1$ chain in a helical magnetic field using the Density Matrix Renormalization Group (DMRG)
\cite{PhysRevLett.69.2863,dmrg} focusing on entanglement measurements and the energy gap from the ground state to low-lying
excitations. Over the past three decades, DMRG has developed into arguably the most powerful
numerical tool for one-dimensional systems \cite{dmrg} and is ideally suited for the present
problem. Both standard DMRG and the infinite-DMRG (iDMRG) were used in the calculations. This allows us a characterization of the phase diagram as a function of pitch angle $\varphi$ and magnetic field strength $B$.

Furthermore, we have developed an analytical theory to obtain a qualitative understanding of the numerically obtained phase diagram. It is known that the ground state properties of a spin-$1$ chain can be conveniently accessed by splitting it into two coupled spins-$1/2$ chains.\cite{SchulzTransfo} The latter spin-$1/2$ systems, in turn, can be mapped onto two coupled spinless fermionic chains via a Jordan-Wigner transformation. The resulting system can be studied using bosonization and a renormalization group analysis.

This paper is organized as follows. First, we present our model in Sec.~\ref{themodel}. In Sec.~\ref{numerics}, we explain the details of the numerical simulations and show the numerically obtained phase diagram of the model, as well as a plot of its entanglement spectrum. In Sec.~\ref{bosonization_section}, the phase transitions will be explained qualitatively based on a bosonization approach. Finally, we present our conclusions in Sec.~\ref{ccl}.

\section{The model}
\label{themodel}

We would like to investigate the phase transition in an isotropic spin-$1$ Heisenberg chain in the presence of a spiral magnetic field. The system is described by the Hamiltonian
\begin{align}\label{HeisenbergHamiltonian}
H  &=J\sum_j \mathbf{S}_j\cdot\mathbf{S}_{j+1}+B\sum_j\left[\cos\left(\varphi j\right)S^x_j+\sin\left(\varphi j\right)S^y_j\right]. 
\end{align}
Here, $\mathbf{S} = (S^x, S^y, S^z)$ and $S^{x,y,z}_j$ are the components of a local spin operator on a site $j$ of a chain of length $L$ and $B$ is the strength of the magnetic field. The latter points in the $x-y$ plane and describes a spiral around the direction of the chain. The angle $\varphi$ becomes the tilt between the orientations of the respective magnetic fields at the sites $j$ and $j+1$. It is already known that this system hosts a Haldane phase in the limit of a constant field ($\varphi = 0$), \cite{pollman_entropy_spectrum} whereas the spectrum is fully gapped and the Haldane phase is absent for a staggered field ($\varphi = \pi)$. Our aim is to investigate the crossover between these two limits.

In order to elucidate the physical relevance of the Hamiltonian (\ref{HeisenbergHamiltonian}), we perform a unitary transformation which corresponds to rotating the spin on site $j$ by an angle $\varphi j$ about the $z$ axis,
\begin{eqnarray}
U=\prod_{j=1}^L U_j \ \ \mathrm{with} \ \ U_j=e^{-i\varphi j S^z_j}.
\label{spin_rotation}
\end{eqnarray}
Under this transformation, Eq.~(\ref{HeisenbergHamiltonian}) is shown to be equivalent to an anisotropic spin-$1$ Heisenberg Hamiltonian with DMI in a constant magnetic field orientated along the $x$-axis,
\begin{align}
H_{DM}&=U H U^{-1}\notag \\
&=\sum_{j=1}^{L-1}\Big[J_x S^x_jS^x_{j+1}+ J_y S^y_jS^y_{j+1} + J_z S^z_jS^z_{j+1} \notag \\
&+\mathbf{D}\cdot\left(\mathbf{S}_j\times\mathbf{S}_{j+1}\right)\Big] +B\sum_j^L S^x_j .
\label{DMHamiltonian}
\end{align}
The resulting DMI vector has the form $\mathbf{D} = (0, 0, D_z)$ and the value of the Heisenberg exchange couplings and the DMI strength for a pitch angle $\varphi$ are given by $J_x = J_y = J \cos(\varphi)$, $J_z = J$, and $D_z = J \sin(\varphi)$. 

\section{Numerical results}
\label{numerics}

The energy gap was estimated from simulations of the Hamiltonian (\ref{HeisenbergHamiltonian}) on a chain of length $L=36$ with periodic boundary conditions. The lowest excited state is obtained by performing a DMRG sweep for the wavefunction of the ground state, then performing another DMRG calculation but with an additional constraint that the second 'ground state' wavefunction has to be orthogonal to the first wavefunction found. The difference in the energies of the two states found is the energy gap $\Delta$. Normally, DMRG works best with open boundary conditions (OBC), and the computational cost of implementing PBC is substantial. However, in this case, the use of PBC is essential. The Haldane chain has a unique ground state for a periodic chain, whereas for an open chain the ground state develops 4-fold degeneracy. As the latter makes the estimation of energy gap numerically very challenging, periodic boundary conditions need to be used. A bond dimension of $m=350$ and 10 DMRG sweeps were found to be sufficient for the energy to converge with a relative error less than $10^{-7}$.

\begin{figure}
   \includegraphics[width=\linewidth,height=6cm]{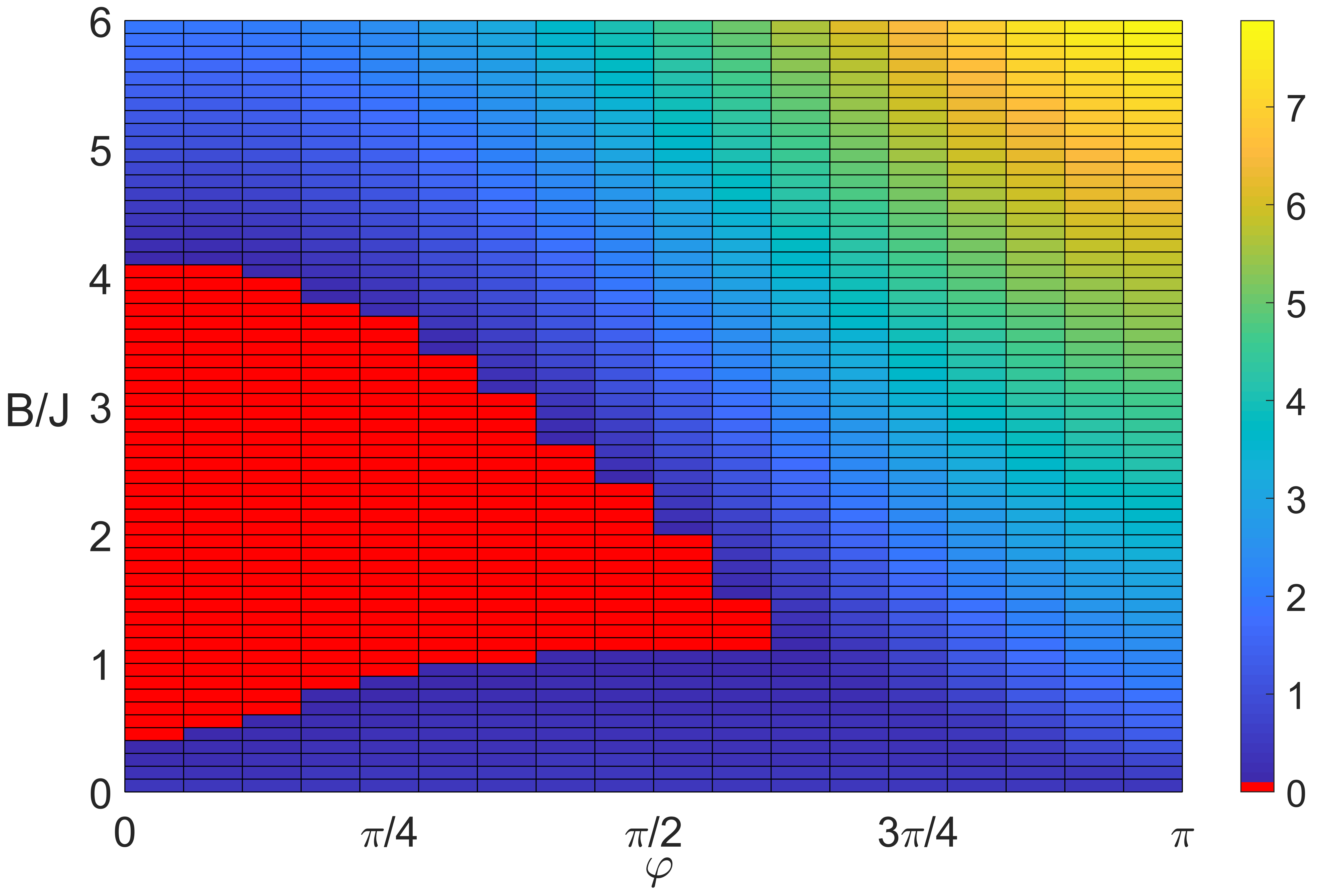}
\caption{Energy gap $\Delta$ of the spin-1 Heisenberg chain in a helical magnetic field as a function of the magnetic field $B$ and of the pitch angle $\varphi$. The red portion represents the gapless region of the energy gap phase diagram. }
\label{phase_diagram}
\end{figure}

The results of the energy gap calculations are shown in Fig.~\ref{phase_diagram}. In the limit of zero magnetic field, the Hamiltonian (\ref{HeisenbergHamiltonian}) reduces to the spin-$1$ Heisenberg model. The ground state in this limit (the Haldane phase)
has a finite gap to the lowest excitations and exhibits no long range magnetic order. However, there is a topological order characterized by the string order parameter that differentiates the state from a quantum paramagnet. Except for this Haldane phase near $B=0$, the onset of the gap in Fig.~\ref{phase_diagram} resembles that of a spin-$1/2$ chain with DMI.\cite{schmidt16} At small, non-zero values of the applied field, the gap remains finite. For values of the twist angle, $\varphi\lesssim 2\pi/3$, the gap vanishes at a critical field, $B_{c1}(\varphi)$
that depends on $\varphi$, accompanied by a transition to a gapless antiferromagnetically ordered ground state. Upon increasing the field further, there is eventually a transition to a fully polarized state at a second critical field, $B_{c2}(\varphi)$ when a (trivial) ground state gap reappears. The field extent of the gapless antiferromagnetic phase
decreases continuously with increasing $\varphi$ and eventually vanishes at
$\varphi\approx 2\pi/3$. For larger twist angles, the ground state of the system evolves from the topologically ordered Haldane phase to a topologically trivial fully polarized phase without any phase transition at any finite non-zero field.

\begin{figure}
   \includegraphics[width=\linewidth,height=6cm]{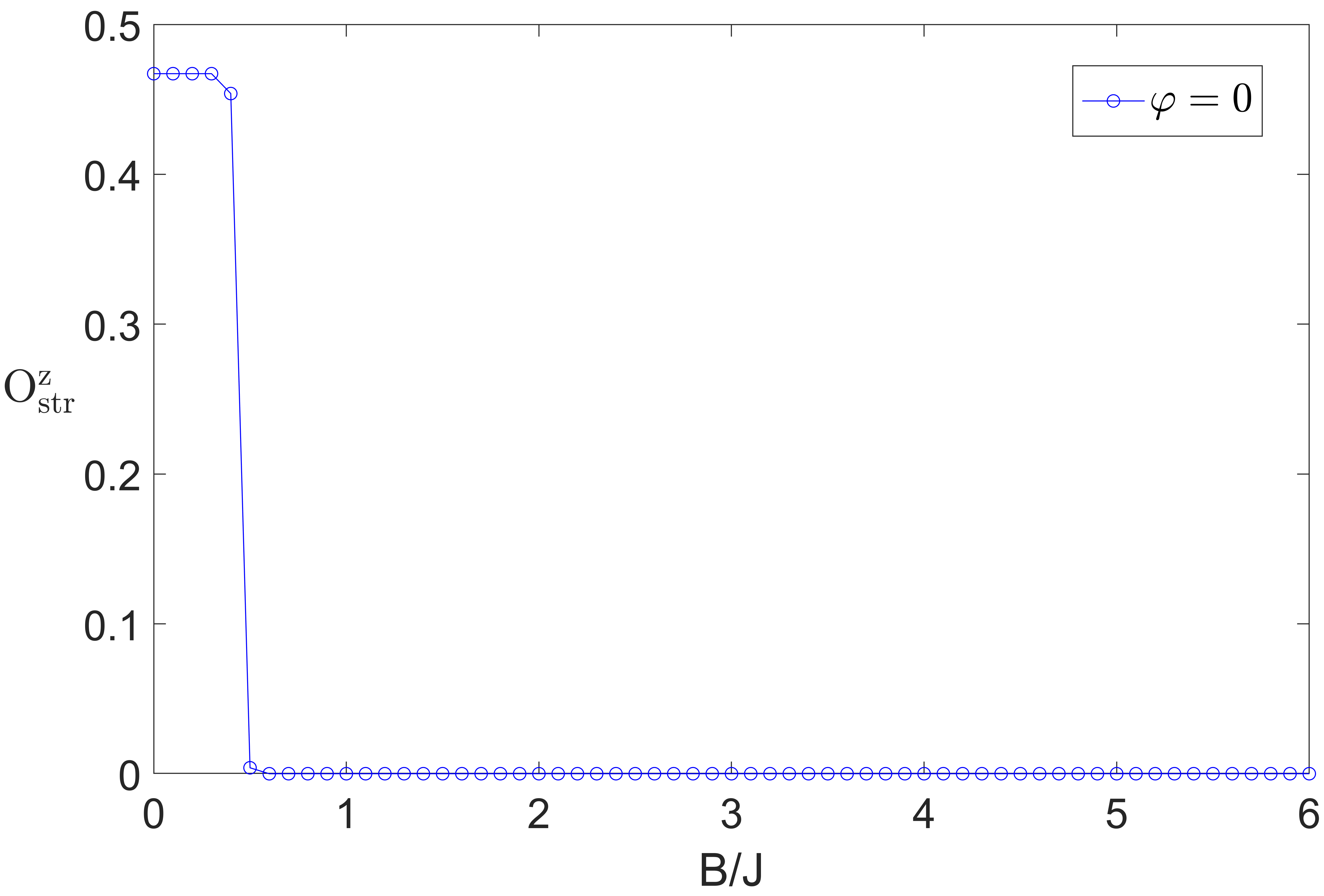}
\caption{String order parameter $O^z_{str}$ at $\varphi=0$. The string order parameter vanishes when the excitation gap closes and the system undergoes a phase transition from the Haldane phase to a topologically trivial one.}
\label{stringorder}
\end{figure}

What is the nature of the ground state in the field range $B < B_{c1}(\varphi)$ for $\varphi \lesssim 2\pi/3$? Is it adiabatically connected to the Haldane phase at $B=0$ and topologically ordered? The string order parameter, defined as
\begin{equation}
O^z_{str}\equiv\lim_{\left|i-j\right|\xrightarrow{}\infty}-\langle S^z_i \left[e^{i\pi\sum^{j-1}_{k=i+1}S^z_k}\right]S_j^z \rangle \ ,
\end{equation}
is a definitive observable to identify the Haldane phase and it shows that increasing a linear magnetic field destroys the Haldane phase (Fig.~\ref{stringorder}). This function also shows that at high magnetic fields where the gap reopens, the resulting spin-polarized phase is topologically different from the Haldane phase.

The $S=1$ Heisenberg chain has global $\mathbb{Z}_2 \times \mathbb{Z}_2$ symmetry,\cite{PhysRevB.45.304,PhysRevB.85.075125} i.e., any $\pi$ rotations about the $x$ and $y$ axes (or equivalently, $y$ and $z$ or $z$ and $x$ axes) transform the state back to itself. However, the introduction of a finite helical field results in the loss of this global symmetry. This makes the string order parameter no longer a valid measure of topological order for $\varphi > 0$. Instead, we have used the structure of the entanglement spectrum to identify the topological character of the different ground state phases. 

\begin{figure}
   \includegraphics[width=\linewidth,height=6.5cm]{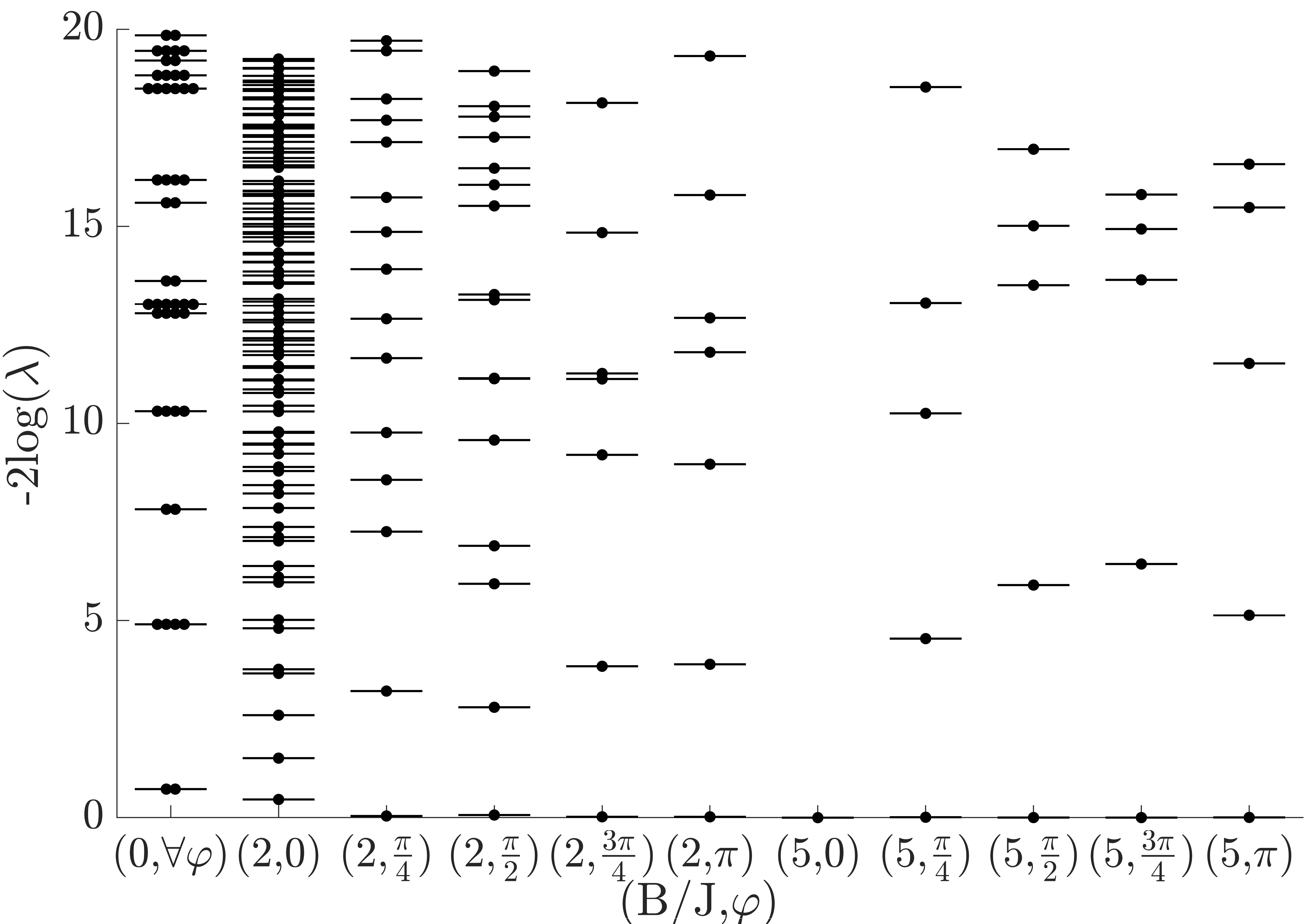}
\caption{Entanglement spectrum for the spin-$1$ Heisenberg chain in a helical magnetic field for various parameters. $\lambda$ are the eigenvalues calculated from the density matrix of the entanglement Hamiltonian. The double degeneracy of the eigenvalues in the Haldane phase is lifted in finite helical magnetic fields.}
\label{eigenspectrum}
\end{figure}

To obtain the entanglement spectrum, one partitions the system into two blocks. The reduced density matrix of either of these blocks can be expressed as $\hat{\rho} = \smash{e^{-\hat{H}}}$. The eigenvalues of the ``entanglement Hamiltonian'' $\hat{H}$ form the entanglement spectrum. For a topologically ordered phase the latter consists of doubly degenerate eigenvalues,\cite{entanglementspec} and it serves as a powerful identifier of the topological character of a state where conventional measures are not applicable. Figure~\ref{eigenspectrum} presents the measured entanglement spectrum for multiple pairs of the parameters $(B/J,\varphi)$ within the different phases. At $B=0$, the system is in the Haldane phase, which is a topologically ordered phase. The entanglement spectrum  exhibits the expected double degeneracy. Results for $\varphi > 0$ show that the double degeneracy is lifted for any non-zero helical magnetic field. At finite helical magnetic fields, there are no topological phases. For $\varphi < 2\pi/3$, the ground state remains gapped, but the topological character is lost. To summarize, the long-range topological character of the Haldane phase is broken at infinitesimally small helical magnetic field. All the other phases (AFM and fully polarized) are topologically trivial as well, as confirmed by their entanglement spectra.

\begin{figure}
   \includegraphics[width=\linewidth,height=6.5cm]{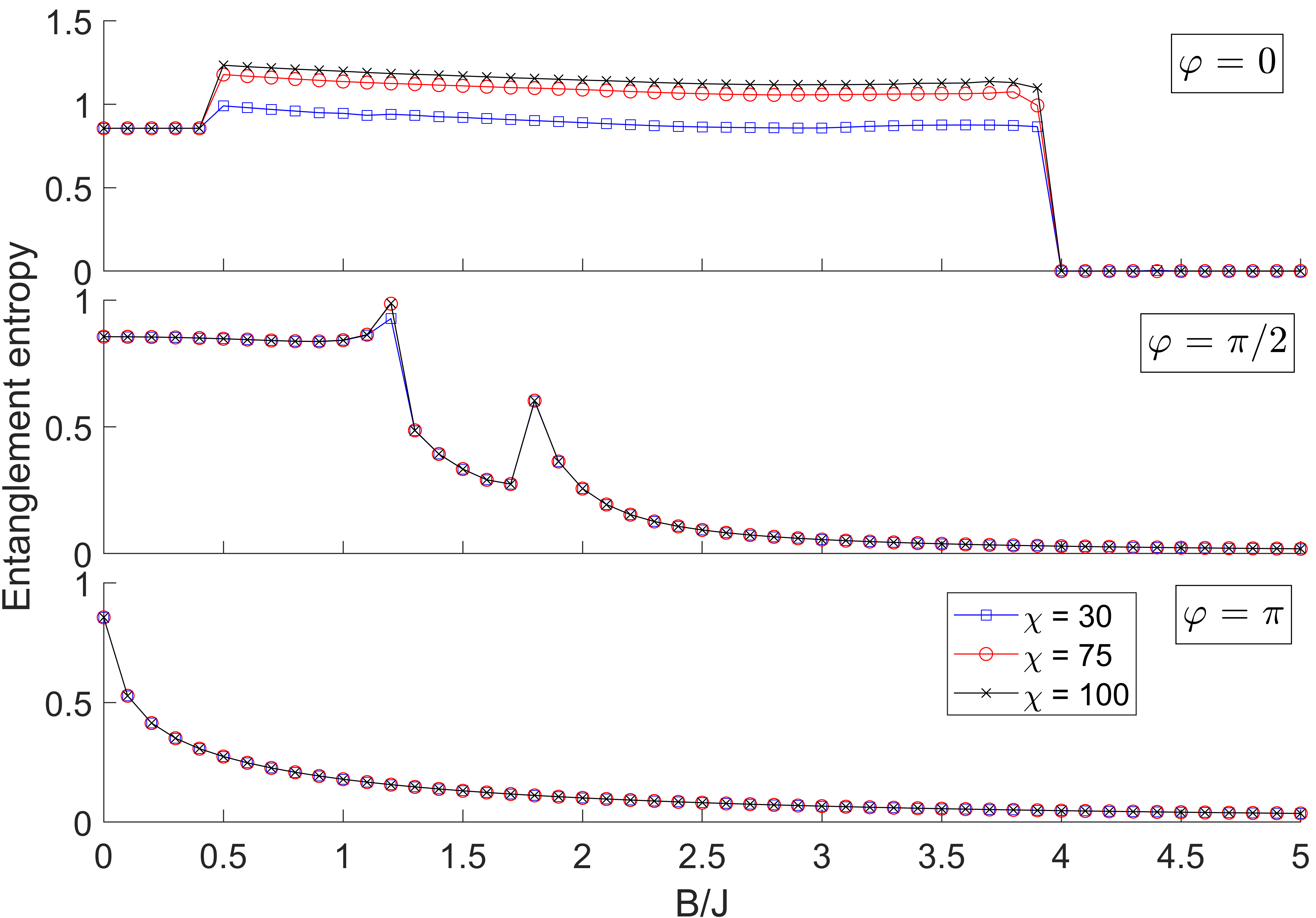}
\caption{Entanglement entropy plots for $\varphi = 0$, $\pi/2$ and $\pi$ for various $\chi$.}
\label{entanglemententropy}
\end{figure}

Finally, the behavior of the entanglement entropy is a reliable indicator of a phase transition. As the entanglement of a system near a phase transition increases without bound, the entanglement entropy is expected to diverge at a critical point. For finite systems, the entanglement entropy should increase with the bond dimension $\chi$, which is the parameter that controls the truncation of the system as iDMRG is performed.\cite{hauschild2018efficient} The results are shown in Fig.~\ref{entanglemententropy} for three representative values of the twist angle. For $\varphi = 0$ and $\pi/2$, the entanglement entropy exhibits singularities at the critical field values for the Haldane-AFM and AFM-fully polarized transitions. On the other hand, for $\varphi = \pi$, the entanglement entropy decreases monotonically with increasing field, reflecting the absence of any transition, thus confirming the phase diagram obtained from energy gap and entanglement spectrum data.

\section{Interpretation of the phase transitions}
\label{bosonization_section}

In this section, we discuss the numerical results by interpreting the phase diagram based on Renormalization Group (RG) arguments. Indeed, to explain the phase transitions presented in Fig.~\ref{phase_diagram}, we start by mapping the Hamiltonian~(\ref{HeisenbergHamiltonian}) onto a ladder of spin-$1/2$ chains, which in turn makes it possible to use a bosonization approach. The ensuing step is to identify the most RG-relevant terms in the bosonized Hamiltonian as a function of the parameters $\varphi$ and $B$.

\subsection{Bosonization of the Hamiltonian}
The different steps to bosonize a spin-$1/2$ Heisenberg Hamiltonian can be found in standard textbooks.\cite{giamarchi_bosonisation} However, because the chain considered here is composed of spin-$1$ sites, the traditional approach is not sufficient to fully bosonize the Hamiltonian  (\ref{HeisenbergHamiltonian}). Instead, we need to use an approach proposed by Schulz,\cite{SchulzTransfo} which rests on the fact that the ground state properties of an $S=1$ Heisenberg Hamiltonian are the same as those of two coupled $S=1/2$ spin chains. Hence, for the consideration of the ground state phase diagram, it is permissible to represent the spin-$1$ operator $\mathbf{S}_j$ on site $j$ in terms of two spins-$1/2$ operators $\mathbf{S}^{1,2}_j$ via the replacement
\begin{align}\label{eq:spinone}
    \mathbf{S}_j =\mathbf{S}^1_j +\mathbf{S}^2_j.
\end{align}

To apply the Jordan-Wigner transformation, which is a prerequisite of bosonization, we proceed to express Eq.~(\ref{HeisenbergHamiltonian}) in terms of Pauli matrices using Eq.~(\ref{eq:spinone}) and $\mathbf{S}^{1,2}_j = \bm{\sigma}^{1,2}_j/2$. Defining the helical magnetic field vector $\mathbf{B}_j = B [\cos(\varphi j), \sin(\varphi j), 0]$, one finds $H = \sum_{r = 1,2} (H_{H,r} + H_{B,r}) + H_{12}$, where
\begin{align}
H_{H,r}&=\frac{J}{4}\sum_j \bm{\sigma}^r_j\cdot\bm{\sigma}^r_{j+1},  \\
H_{B,r} &=\frac{B}{2}\sum_j\Big[(\cos\left(\varphi j\right)\sigma^{x,r}_j
 +\sin\left(\varphi j\right)\sigma^{y,r}_j\Big], \\
H_{12}&=\frac{J}{4}\sum_j \left(\bm{\sigma}^1_j\cdot\bm{\sigma}^2_{j+1}+\bm{\sigma}^2_j\cdot\bm{\sigma}^1_{j+1}\right).
\end{align}
The system thus consists of two identical spin-$1/2$ chains, each of which interacts with a spiral magnetic field. Using the language of spin ladders, \cite{ueda_dmi_spinladder} these chains are coupled by a cross inter-chain coupling term $H_{12}$, where every site $j$ of one leg interacts with the site $j+1$ of the other leg. At low energies, the Heisenberg terms $H_{H,r}$ can be expressed in terms of the bosonic fields $\phi\left(x\right)$ and $\theta\left(x\right)$ as
\begin{align}
    H_{H,r}
&=
    \int \frac{dx}{2\pi} \left[\frac{u}{K}\left(\nabla\phi^r\left(x\right)\right)^2+uK\left(\nabla\theta^r\left(x\right)\right)^2\right] \notag \\
&-
    \frac{J}{2a\pi^2}\int \ dx\cos\left(4\phi^r\left(x\right)\right),
\label{LuttingerHamiltonian}
\end{align}
where $a$ is a cutoff related to the lattice spacing, $u/K$ is a parameter directly related to the compressibility of the system and $K$ is the Luttinger parameter.

Next, to express the coupling term between the two chains with a bosonic representation, we use the representation,\cite{giamarchi_bosonisation,SchulzTransfo} 
\begin{align}
\sigma^\pm\left(x\right)& \propto e^{\mp i\theta\left(x\right)}\left[\left(-1\right)^x+\cos\left(2\phi\left(x\right)\right)\right],
\label{sigmapm}
\end{align}
which leads to

\begin{align}
    H_{12}
&\propto
    J\int dx \Big[\cos\left(\theta^2(x) - \theta^1(x)\right)\notag\\
&+
    \sum_{j=0}^1 4\cos\left(2\phi^1(x) + \left(-1\right)^{j} 2\phi^2(x) \right)\Big].
\end{align}

\begin{figure}
   \includegraphics[scale=0.3]{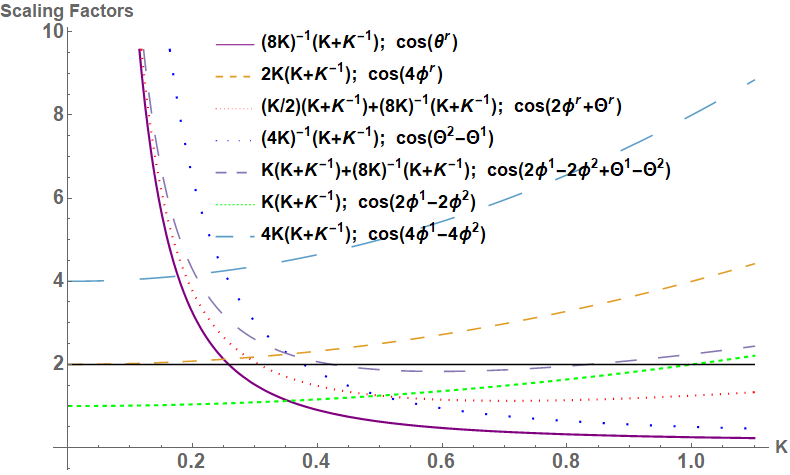}
\caption{Scaling dimensions associated with the different cosines present in $H$ as a function of $K$. The full lines represent cosines that are the most relevant for Luttinger parameters $K > 0.4$. The black line is the limit above which any cosine becomes irrelevant.}
\label{scaling_factors}
\end{figure}

Last but not least, for the magnetic field term we need to use Eq.~(\ref{sigmapm}) again, and obtain
\begin{align}
H_{B,r}&\propto B\int dx\left[ \left(-1\right)^x\right.
\nonumber\\&\left.+\cos\left(2\phi^r\left(x\right)\right)\right]\cos\left(\varphi x+\theta^r\left(x\right)\right).
\label{ZeemanHamiltonian}
\end{align}
For determining the scaling dimensions, it is convenient to remove $\varphi$ from the cosine terms via the change of variable to $\Theta^r(x)=\theta^r(x)+\varphi x$. Moreover, one can neglect fast oscillating terms proportional to $(-1)^{x}$. The calculations of the scaling dimensions of the various cosine terms is then straightforward and the result is shown in Fig.~\ref{scaling_factors}.

\subsection{The different phases}
\label{diferrent_phases}

A single RG-relevant cosine term depending on the fields $\phi$ or $\theta$ in a bosonized Hamiltonian will in general open a gap in the spectrum. Consider as an example the term $-\cos(4 \phi(x))$. If it is relevant, the ground state is reached by pinning $\phi(x)$ to a minimum of the cosine term, e.g., $\phi(x) = 0$. A quadratic expansion around $\phi(x) = 0$ then reveals that the spectrum becomes gapped. 

However, in our discussion, the picture is more complex for two reasons: firstly, an RG-relevant cosine term only becomes the dominant term in the Hamiltonian if the RG-flow can be continued all the way to zero energy. But if another term with a large energy scale is present, the RG flow stops at that energy scale, in which case the most RG-relevant term is not necessarily the dominant term in the Hamiltonian. Secondly, the situation is more complex in the presence of competing cosine terms. If multiple non-commuting cosine terms are present, it is impossible to pin non-commuting phase operators separately to their respective minima. In that case, the competition between two relevant cosine terms may cause the system to remain gapless. Hence, to find the phase of the system for a given choice of parameters, we have to take into account both the relevance of the cosine terms as well as the value of their prefactors.

As the Luttinger parameter $K$ depends on the coupling $J$, the RG relevance of the cosines will 
change along the vertical axis of the phase diagram in Fig.~\ref{phase_diagram}. The phase 
transitions along this axis are thus determined by the changing order of relevance of different 
cosine terms. In contrast, along the horizontal axis, only the parameter $\varphi$ changes, not the scaling dimensions, which means that only the relative importance of the prefactors as a function of $\varphi$ and $J$ 
determines the phase transition.

The gapped Haldane phase only appears in integer spin chains. Therefore, the origin of the gap
opening must be found in $H_{12}$. It turns out that, when considering the limit $B=0$, we do find 
relevant cosines containing different combinations of the fields $\theta^{r}$ and $\phi^{r}$ in 
$H_{12}$. It is the presence of at least one of these cosines that opens a gap, explaining thus the 
gapped phase for small values of the ratio $B/J$. The bosonized expression of the string order 
parameter contains $\cos(\phi^r(x))$ and $\nabla\phi^r(x)$ coming from the bosonized expression of 
$S_j^z(x)$.\cite{giamarchi_bosonisation} When the term $\cos(2\phi^r(x))$ is relevant, which 
corresponds to low values of $K$, fluctuations of $\phi$ are suppressed and we find indeed 
$O^z_{str}\approx 0$. According to Fig.~\ref{stringorder}, the string order parameter vanishes 
towards large values of $B/J$. Therefore, we can deduce that $K<1/2$ corresponds to the large values 
of $B/J$, while $K>1/2$ corresponds to small values of $B/J$. Eventually, it implies that 
$\cos(\Theta^{2}-\Theta^{1})$ is the most relevant cosine term in the Haldane phase because it is the 
most relevant term for $K>1/2$.

The phase for $B/J \gg 1$ is also a gapped phase. In this case, the large prefactor $B/J$ means that one can neglect every cosine term except the magnetic field term $\cos(2\phi^r(x)\pm\Theta^r(x))$. These cosine terms are therefore at the origin of the gap opening in this regime, and the resulting phase corresponds to a trivial spin-polarized phase. Note that in this case the magnetic field term is dominant despite the fact that it is not the most RG-relevant term, simply because the RG-flow can only be continued up to energies of order $B$. 

We have identified the dominant cosine terms for the extreme values of $B/J$. We will now study more in detail the two phase transitions occurring at $B/J\approx 0.4$ and $B/J \approx 4$, both at $\varphi=0$. For $0.4<B/J< 4$, the prefactors of different cosine terms of $H$ are of same order. As we argued before, upon increasing $B/J$ the value of the Luttinger parameter varies from $K>1/2$ to $K<1/2$ where this time $\cos(2\phi^1(x)\pm 2\phi^2(x))$ is the most relevant. Additionally, for $K=1/2$ the system reaches a certain point where all the three cosines of interest are equally relevant. It can be deduced that the gapped to gapless phase transition occurring at $B/J=0.5$  corresponds to the moment when $\cos(2\phi^1(x)\pm 2\phi^2(x))$ and $\cos(2\phi^r(x)\pm\Theta^r(x))$ become as relevant as $\cos(\Theta^1(x)\pm\Theta^2(x))$. Then, the phase transition at $B/J=4$ logically occurs because $\cos(2\phi^1(x)\pm 2\phi^2(x))$ becomes the most relevant of the cosines. 

We can isolate the term in the Hamiltonian corresponding to the helical magnetic field,
\begin{eqnarray}
H_{\varphi}=- 2 u K\varphi \sum_{r=1}^2 \int dx \nabla \Theta^r(x).
\end{eqnarray}
Therefore, in bosonization language this term is equivalent to a chemical potential term for both spin-$1/2$ chains. This allows us to understand the phase transition which occurs upon increasing $\varphi$ for a given $B/J$ as analogous to a Mott transition from a gapless metallic phase to a gapped Mott insulating phase due to a change of chemical potential, i.e., a so called Mott-$\delta$ phase transition.\cite{giamarchi_bosonisation}

The two previous phenomena are helpful in understanding the approximately triangular shape of the gapless region in the phase diagram. Indeed, since increasing both $B$ or $\varphi$ leads to a trivial gapped phase characterized by the same spin-polarized state, the transition to the gapped phase occurs for larger (smaller) $\varphi$ as $B/J$ is decreased (increased). This explains qualitatively the fact that the line separating the metallic phase from the insulator phase in the phase diagram has a negative slope.

Finally, a comment is due about the limit of a staggered magnetic field corresponding to $\varphi=\pi$. In this case the magnetic field term becomes
\begin{eqnarray}
H_{B1}+H_{B2}=\sum_{r=1}^2 \frac{B}{2\pi a}\int dx \ \cos\left(\theta^r(x)\right).
\end{eqnarray}
For $0.35<K<1$ this cosine term becomes the most RG relevant, see Fig.\ref{scaling_factors}. Its prefactor depends on the values of $B$, therefore for $0.35<K<1$ and a strong enough magnetic field, this term will be at the origin of the gap opening. For large values of the magnetic field corresponding to $K<0.35$, although $\cos\left(\theta^r(x)\right)$ is not the most relevant cosine anymore, the other cosine terms are negligible compared to it and, consequently, $\cos\left(\theta^r(x)\right)$ will still be the cosine opening the gap. In contrast, for small values of $B$, in other words large values of $K$, this term can be neglected compared to $\cos(\Theta^{2}-\Theta^{1})$ which is for these values of $K$ the second most RG relevant cosine and will be thus at the origin of the gap opening. However, these two terms commute, so there is no gap closing between two different regimes as $B$ is increased.

\section{Conclusion}
In summary, we have studied the ground state phases of the $S=1$
Heisenberg chain with DM interactions in a  magnetic field 
using complementary numerical and analytical techniques. The ground state of the model in the
absence of DMI and external field is the Haldane phase which is
a symmetry-protected topological (SPT) phase protected by a
$\mathbb{Z}_2 \times \mathbb{Z}_2$ symmetry. The interplay between DMI and a magnetic field leads to an effective spiral magnetic 
field which breaks this symmetry. Consequently, the ground state of 
the present model loses the SPT character for {\it any} non-zero DMI 
or equivalently, spiral magnetic field. We have identified the various field-driven
phases and the associated phase transitions and studied their 
topological character using the entanglement spectrum. 
The phase transitions obtained numerically have been analytically confirmed by a bosonization study.
\label{ccl}

\section*{Acknowledgements}

HT and TLS acknowledge support from the Fonds National de la Recherche
Luxembourg under the grants AFR 11224060 and ATTRACT 7556175.
ETSO and PS acknowledge support from the Ministry of Education, Singapore,
under the grant MOE2016-T2-1-065.
ETSO and PS would like to thank F. Pollmann and J. Hauschild for the insightful discussions as well as the HPC Team, High Performance Computing Centre, NTU for their expertise and help rendered.

\bibliography{bibliography}

\end{document}